\documentclass[12pt]{easychair}

\newcommand{\xe}[1]{Equation~\ref{#1}}

\newcommand{\gipsy}{{GIPSY\index{GIPSY}}}

\newcommand{\lucid}{{Lucid\index{Lucid}}}

\newcommand{\flucid}{{Forensic Lucid\index{Forensic Lucid}}}

\newcommand{\trans}{$\psi$}

\newcommand{\invtrans}{$\Psi^{-1}$}

\newcommand{\dmarf}[0]{DMARF\index{MARF!Distributed}\index{Frameworks!Distributed MARF}\index{Libraries!Distributed MARF}}

\newcommand{\lucidL}[1]{{$\mathit{Lucid}$}($L$) }

		{}

\def\myvert{\raise 2.27pt \hbox{\vrule depth 0pt height 8pt width 0.2mm}}
\def\myarrow{\hspace*{0.43mm}%
             \raise 2.29pt\hbox{\vrule depth 0pt height 8pt width 0.16mm}%
             \hspace*{-0.32mm}%
             $\longrightarrow$
             \ %
             }

\newcommand{\assl}{ASSL\index{ASSL}}
\newcommand{\smart}{S.M.A.R.T.}

\begin{document}

\title{Towards Improving Validation, Verification,
       Crash Investigations, and Event Reconstruction
       of Flight-Critical Systems with Self-Forensics}

\titlerunning{Self-Forensics for Flight-Critical Systems, RFI NNH09ZEA001L}

\author{Serguei A. Mokhov\\
Department of Computer Science and Software Engineering\\
Concordia University, Montreal, Canada\\
Tel: +1 (514) 848-2424 x5179\\
Fax: +1 (514) 848-3299\\
\url{mokhov@cse.concordia.ca}\\
}

\authorrunning{Mokhov}

\maketitle

\begin{abstract}
This paper introduces
a novel concept of {\em self-forensics}
to complement the standard autonomic self-CHOP properties of the self-managed systems,
to be specified in the {\flucid} language. We argue that self-forensics, with the
forensics taken out of the cybercrime domain, is applicable to ``self-dissection''
for the purpose of verification of autonomous software and hardware systems of flight-critical systems
for automated incident and anomaly analysis and event reconstruction by
the engineering teams in a variety of incident scenarios during design and
testing as well as actual flight data.
\end{abstract}

\section{Introduction}

In this paper we introduce a new concept for flight-critical
integrated software and hardware systems to analyze themselves
forensically as needed as well as keeping forensics data
for further automated analysis in cases
of reports of anomalies, failures, and crashes.
We insist this should be a part of the protocol for
each system, (even not only flight systems), but any large and/or critical
self-managed system.

This proposition is a rehash of the related work of the author during his PhD
studies~\cite{self-forensics-flucid-as,self-forensics-flucid-road-vehicles}
for the NASA spacecraft self-forensics concept as well as a work towards
improving the safety and crash investigation of read vehicles with similar
means.

We review some of the related work that these ideas are built
upon prior describing the requirements for self-forensics components.
We describe the general requirements as well as limitations and
advantages.
This is a draft sketch.

\subsection{Applicability Overview and Discussion}

Many ideas in this work come from computer forensics and forensic computing.
Computer forensics has traditionally been associated with
computer crime investigations. We show the approach
is useful as an aid in for validation and verification during design, testing, and simulations
of aircraft systems as well as during the actual in-flight operations and
crash investigations. We earlier argued~\cite{self-forensics-flucid-as,self-forensics-flucid-road-vehicles}
if the new technologies
are built with the self-forensics components, it would
even help space and automotive industries or anything robotic
and autonomous, and has large complex software systems,
including military.

Existing self-diagnostics,
computer BIOS reports,
{\smart}~\cite{wiki:SMART}
reporting for hard disk as well as many other
devices could be a good source for such data
computing, i.e. be more forensics-friendly
and provide forensics interfaces for
self-forensics analysis and investigation
as well as allowing engineering teams extracting,
analyzing, and reconstructing events using such data.

Thus, we insist that self-forensics, if included
earlier in the design and development of the spacecraft,
not only helps during the validation and verification,
but also a posteriori, during the day-to-day operations
of the airborne and spaceborne systems.

Some example cases where self-forensics would have been helpful
to analyze anomalies say in the spacecraft, when Mars Exploration Rovers behave
strangely~\cite{nasa-mer-20090128-blankout},
or even with one is doing a hard disk
recovery, such as from the shuttle Columbia~\cite{harddisk-recovery-columbia},
or automatically as well as interactively reasoning about events,
possibly speeding up the analysis of the anomalies in subsystems.
Another example is when the Hubble Space Telescope was switched from
the side A of its instruments to the redundant side B. The self-forensics units would
help Hubble to analyze the problem and self-heal later.
Of course, the cost of such self-forensic units would not be negligible;
however, the cost of self-forensics units maybe well under
than the costs of postponing missions, as e.g. happening
with the Hubble Space Telescope Servicing Mission 4 (SM4)
and the corresponding shuttle processing delay and costs
of moving shuttles around%
~\cite{nasa-hubble-20081017-status-3,%
nasa-hubble-20081017-status-4,%
nasa-hubble-20081204-may-launch,%
nasa-hubble-20080929-telecon,%
nasa-hubble-20081230-status,%
nasa-hubble-20081219-status} and others.

Further more, the concept of self-forensics would
be even a greater enhancement and help with flight-critical systems,
the blackboxes in aircraft, etc. to help with crash
investigations~\cite{scotland-helicopter-crash-failure}.

\subsection{Self-Management Properties}

The common aspects of self-managing systems, such
as self-healing, self-protection, self-optimization,
and the like (self-CHOP) are now fairly well understood
in the literature and R\&D~\cite{ibmarchblprnt2006,kephartacvis03,truszkowski04,hinchey05,assl-nasa-swarm-sac08,autonomic-computing-2006,autonomic-computing-2004}.
We augment that list with {\em self-forensics} that we would
like to be a part of the standard list of autonomous
systems specification.

The self-forensics property is meant to embody and formalize
all existing and future aspects of self-analysis, self-diagnostics,
data collection and storage, software and hardware components (``sensors'')
and decision making that were
not formalized as such and define a well-established
category in the industry and academia.
In that view, self-forensics encompasses self-diagnostics, blackbox recording,
(Self-Monitoring, Analysis, and Reporting Technology) {\smart} reporting~\cite{wiki:SMART}, and encoding
this information in analyzable form of {\flucid} (or some
other format if desired when the concept matures) for later automated analysis and
even reconstruction using the corresponding expert system tool.
Optional parallel logging of the forensics events during
the normal operation of the aircraft, especially
during the blackout periods will further enhance the durability
of the live forensics data logged from the spacecraft to the
nearby control towers or flight centers.

\subsection{{\flucid}}

{\flucid}~\cite{flucid-imf08,flucid-isabelle-techrep-tphols08,marf-into-flucid-cisse08}
is a forensic case specification language for automatic deduction and event reconstruction
of computer crime incidents. The language itself is general enough to specify any events,
their properties, duration, as well as the context-aware system model. We take out
the {\flucid} from the cybercrime context for application to any autonomous software
or hardware systems as an example of self-forensic case specification.

{\flucid} is based on
{\lucid}~\cite{lucid85,lucid95,lucid76,lucid77,eager-translucid-secasa08}
and its various dialects that allow
natural expression of various phenomena, inherently parallel, and
most importantly, context-aware, i.e. the notion of context 
is specified as a first-class value in
{\lucid}~\cite{gipsy-simple-context-calculus-08,wanphd06}.
{\lucid} dialects are functional programming languages.
All these properties make {\flucid} an interesting choice
for forensic computing in self-managed systems to complement
the existing self-CHOP properties.

{\flucid} is also significantly influenced by and is meant
to be a usable improvement of the work of Gladyshev et al. on
formal forensic analysis and event reconstruction using
finite state automate (FSA) to model incidents and reason
about them~\cite{blackmail-case,printer-case}.

While {\flucid} itself is still being finalized as a part of
the PhD work of the author along with its compiler,
run-time and development environments, it is well under way
to validate its applicability to various use-cases and
scenarios.

\subsubsection*{Context}

{\flucid} is context-oriented.
The basic context entities comprise an observation $o$
in \xe{eq:o}, observation sequence $os$ in \xe{eq:os},
and the evidential statement in \xe{eq:es}. These
terms are inherited from~\cite{blackmail-case,printer-case}
and represent the context of evaluation in {\flucid}.
An observation of a property $P$ has a duration between
$[\min,\min+\max]$. This was the original definition of $o$
and the author later added $w$ to amend each observation
with weight factor or probability or credibility to
later further model in accordance with the mathematical
theory of evidence~\cite{shafer-evidence-theory}.
$t$ is an optional timestamp as in a forensic log for that property.
An observation sequence represents, which is
a chronologically ordered collection of observations
represent a story witnessed by someone or something
or encodes a description of some evidence. All these
stories (observation sequences, or logs, if you will)
all together represent an evidential statement about
an incident.
The evidential statement is an unordered
collection of observation sequences. The property
$P$ itself can encode anything of interest~--~an element
of any data type or even another {\flucid} expression,
or an object instance hierarchy or an event.

\begin{equation}
o = (P,t,\min,\max,w)
\label{eq:o}
\end{equation}

\begin{equation}
os = \{o_1,\ldots,o_n\}
\label{eq:os}
\end{equation}

\begin{equation}
es = \{os_1,\ldots,os_m\}
\label{eq:es}
\end{equation}

Having constructed the context, one needs to built
a {\em transition function} {\trans} and its inverse {\invtrans}.
The generic versions of them are provided by {\flucid}~\cite{flucid-isabelle-techrep-tphols08}
based on~\cite{printer-case,blackmail-case}, but the investigation-specific
one has to be built, potentially visually, by the engineering
team, which can be done even before the system, such as spacecraft,
launches, if the self-forensics aspect is included into the design
from the start. The specific {\invtrans} takes evidential statement
as an argument and the generic one takes the specific one.

\subsubsection*{Advantages}

Context-aware, built upon the intensional logic and the {\lucid} language
that existed in the literature and math for more then
30 years and served initial purpose of program verification~\cite{lucid76,lucid77}.

\section{Self-Forensics Application and Requirements}

In this section we elaborate in some detail on the application
of self-forensics and its requirements that must be formal
and included in the design of the flight systems early on.

\subsection{Application}

There are usually some instruments and sensors on board
of many aircraft and airborne vehicles these days,
including high-end computers.
All of them can also have additional functional units to observe
other instruments and components, both hardware and software,
for anomalies and log them appropriately for forensics purposes.

Such forensic specification is also useful to train new engineers on a team,
flight controllers, and others involved, in data analysis, to avoid
potentially overlooking data and making incorrect ad-hoc 
decisions. In an {\flucid}-based expert system (that's what
was the original purpose of {\flucid} in the first place
in cybercrime investigations) one can accumulate a number
of contextual facts from the self-forensic evidence and
the trainees can construct their theories of what happened
and see of their theories agree with the evidential data.
Overtime, (unlike in most cybercrime investigates) it can accumulate the
general enough contextual knowledge base of encoded facts
that can be analyzed across flights, missions, globally
and on the web when multiple agencies and aircraft manufactures
collaborate.

\subsection{Requirements}

Here we define the general requirements scope for the autonomic
self-forensics property adapted for airborne vehicles:

\begin{itemize}
\item
Must always be included in the design and specification.

\item
Should be optional if constrained by the severe budget cuts
for less critical flight components.
Must not be optional for mission critical and
safety-critical systems and blackboxes.

\item
Must cover all the self-diagnostics events, e.g.
{\smart}-like capabilities and others.

\item
Must have a formal specification
(that what it makes it different from just self-diagnostics).

\item
Must have tools for automated reasoning and
reporting about incident analysis matching the specification,
real-time or a posteriori during investigations.

\item
Context should be specified in the terms of system
specification involving the incidents, e.g. parts
and the software and hardware engineering design specification
should be formally encoded (e.g. in {\flucid}) during
the design and manufacturing. This are the static forensic
data. The dynamic flight forensics data are recorded in real-time
during the vehicle operation.

\item
Preservation of forensic evidence must be atomic, reliable,
robust, and durable.

\item
The forensic data must be able to include any or all related
non-forensic data for analysis when needed, e.g. reconnaissance or
science images for military, exploration, and scientific aircraft
taken by a camera around the time of incident or
measurements done around the incident by an instrument
or even the entire trace of a lifetime of a system
logged {\em somewhere} for automated analysis and event
reconstruction.

\item
Levels of forensic logging and detail should be
optionally configurable in collaboration with
other design requirements in order not to hog
other activities resources, create significant overhead,
or fill in the bandwidth of downlinks or to preserve
power.

\item
Self-forensics components should optionally be
duplicated in case themselves also fail.

\item
Event co-relation optionally should be specified.

\item
Some forensic analysis can be automatically
done by the autonomous system itself (provided having
enough resources to do so), e.g.
when it cannot communicate with flight controllers.

\end{itemize}

\subsection{Limitations}

The self-forensics autonomic property is very good
to have for automated analysis of simulated (testing and verification)
and real incidents in autonomous hardware and software systems in aircraft,
but it can not be mandated as absolutely required
due to a number of limitations it creates. However, whenever
the monetary and time budgets allow, it should be included
in the design and development of the autonomous spacecraft, military
equipment, or software systems.

\begin{itemize}

\item
The cost of the overall aircraft systems will obviously increase.

\item
If built into software, the design and development
requires functional long-term storage and CPU power.

\item
It will likely increase of bandwidth requirements; e.g. for scientific and
exploratory aircraft if the science data is doubled in the forensic stream.
If the forensic data are mirrored into the scientific one,
more than twice bandwidth and storage used.

\item
An overhead overall if collect forensics data
continuously. Can be offloaded along with the
usual science and control data down to the
flight control towers or centers periodically.

\item
The self-forensics logging and analyzing software
ideally should be in ROM or similar durable flash type of
memory; but should allow firmware and software
upgrades.

\item
We do not tackle other autonomic requirements of
the system assuming their complete coverage and
presence in the system from the earlier developments
and literature, such as self-healing, protection,
etc.

\item
Transition function has to be modeled by the 
engineering team throughout the design phase
and encoded in {\flucid}. Luckily the data-flow graph (DFG)~\cite{yimin04}
interactive development environment (IDE), like a CAD application
is to be available.

\end{itemize}

\subsection{Brief Example}

\begin{itemize}
\item
	self-forensic sensors observe subsystems and instruments and
	of an aircraft in flight
	
\item
	every engineering or scientific event
	is logged using {\flucid}
	
\item
	each forensic sensor may observes several subsystems or instruments
	
\item
	each sensor ``composes'' a ``story'' in the for of {\flucid} observational
	sequence about an subsystem or instrument
	
\item
	a collection of ``stories'' from multiple sensors, properly encoded,
	represent the evidential statement, either during the verififcation
	or actual flight
	
\item
	if an incident (simulated or real) happens; engineers define theories about what happened.
	The theories are encoded as observation sequences. When
	evaluated against the collected forensic evidence.
	Then the evaluating software system (e.g. in the case of author's PhD work is the General
	Intensional Programming system or {\gipsy})
	can automatically verify the theory matched up against
	the context of evidential statement and if the theory $T$ agrees
	with the evidence, meaning this theory has an explanation
	within the given evidence (and the amount of evidence can
	be significantly large for ``eyeballing'' it by humans),
	then likely the theory is a possible explanation of what
	has happened. It is possible to have multiple explanations
	and multiple theories agreeing with the evidence. In the
	latter case usually the ``longer'' (in the amount of
	events and observations involved) theory is preferred
	or the one that has a higher cumulative weight/credibility
	$w$. Given the previously collected and accumulated knowledge base
	of {\flucid} facts, some of the analysis and events reconstructed
	can be done automatically.
\end{itemize}

\section{Conclusion and Future Work}

We introduced a novel concept of self-forensics with {\flucid}
to aid validation and verification of flight-critical systems
during the design, manufacturing, and testing, as well as its
continuous use during the actual flight and operation of airborne
vehicles.

We drafted some of the requirements for such property to be
included into the design as well as its potential most prolific
limitations today.

We argued that such a property of self-forensics formally grouping,
self-monitoring, self-analyzing, self-diagnosing systems along with
a decision engine and a common forensics logging data format can
standardize the design and development not only airborne vehicles,
but also road vehicles and spacecraft, or even marine and submarine
vehicles, to improve safety and autonomicity of the ever-increasing complexity of the software and
hardware systems in such vehicles and further their analysis
when incidents happen.

We outlined some of the background work, including the forensic
case specification language, {\flucid}, that we adapted from the
cybercrime investigations domain to aid the validation and verification
of the aircraft subsystems design by proposing logging the forensics
data in the {\flucid} context format available for manual/interactive
analysis on the ground as well as real-time by a corresponding
expert system

The generality of the approach manifests itself not only for
the design, manufacturing, development, and testing of the flight
components, as well as system's normal operation once deployed,
but as well as training of the engineering and flight controller
personnel in investigation techniques with a help of a {\flucid}-based
expert system as well as common format for sharing the data and collaboration
between various agencies, such as NASA, and flight hardware and software
manufacturers to improve overall safety of modern- and future-day vehicles. 

Some of the author's future work will include the below, which is
probably less important for the RFI, but nonetheless planned to
be carried out:

\begin{itemize}
\item
Amend {\assl}~\cite{assl-sew,assl-seams,vassevicsoft06,vassevPhDThesis} to handle the self-forensic
property.

\item
Implement the notion of self-forensics in the
{\gipsy}~\cite{agipsy-07,gipsy2005,gipsy-multi-tier-secasa09,gipsy-hoil}
and
{\dmarf}~\cite{dmarf-web-services-cisse08,dmarf-assl-self-healing,dmarf-assl-self-protection,dmarf-assl-self-optimization}
systems the author is closely working on.
\item
Finalize implementation of the {\flucid} compiler
and the development and run-time environment.
\item
Implement large realistic cases encoded in {\flucid}
to test and validate various aspects of correctness,
performance, and usability.
\end{itemize}

\bibliographystyle{unsrt}
\bibliography{self-forensics-nasa-rfi-NNH09ZEA001L}

\end{document}